# Type-II Ising superconductivity and anomalous metallic state in macro-size ambient-stable ultrathin crystalline films

Yi Liu[†,Δ,‡], Yong Xu[Δ,§,∥,‡], Jian Sun[†,Δ,‡], Chong Liu[§,‡], Yanzhao Liu[†,Δ], Chong Wang[Δ,§,⊥], Zetao Zhang[Δ,§], Kaiyuan Gu[†,Δ], Yue Tang[†,Δ], Cui Ding[§], Haiwen Liu[#], Hong Yao[Δ,§,⊥], Xi Lin[†,Δ,∥,¶,\*], Lili Wang[Δ,§,\*], Qi-Kun Xue[Δ,§,∥] & Jian Wang[†,Δ,§,∥,¶,\*]

[†]International Center for Quantum Materials, School of Physics, Peking University, Beijing 100871, China.

[Δ]Collaborative Innovation Center of Quantum Matter, Beijing 100871, China.

[§]State Key Laboratory of Low-Dimensional Quantum Physics, Department of Physics, Tsinghua University, Beijing 100084, China.

[∥]RIKEN Center for Emergent Matter Science, Wako, Saitama 351-0198, Japan

[⊥]Institute for Advanced Study, Tsinghua University, Beijing 100084, China.

[#]Center for Advanced Quantum Studies, Department of Physics, Beijing Normal University, Beijing 100875, China.

[∥]Beijing Academy of Quantum Information Sciences, Beijing 100193, China.

[¶]CAS Center for Excellence in Topological Quantum Computation, University of Chinese Academy of Sciences, Beijing 100190, China.

ABSTRACT: Recent emergence of two-dimensional (2D) crystalline superconductors has provided a promising platform to investigate novel quantum physics and potential applications. To reveal essential quantum phenomena therein, ultralow temperature transport investigation on high quality ultrathin superconducting films is critically required, although it has been quite challenging experimentally. Here we report a systematic transport study on the ultrathin crystalline $PdTe_2$ films grown by molecular beam epitaxy (MBE). Interestingly, a new type of Ising superconductivity in 2D centrosymmetric materials is revealed by the detection of large in-plane critical field more than 7 times Pauli limit. Remarkably, in perpendicular magnetic field, we provide solid evidence of anomalous metallic state characterized by the resistance saturation at low temperatures with high quality filters. The robust superconductivity with intriguing quantum phenomena in the macro-size ambient-stable ultrathin $PdTe_2$ films remains almost the same for 20 months, showing great potentials in electronic and spintronic applications.

Recent developments of film-growth and device-fabrication techniques have triggered a flurry of intensive investigations on 2D crystalline superconducting systems, where new quantum phases and quantum phase transitions are attracting tremendous interest in condensed matter physics[1-10]. The superconductor-insulator/metal transition (SIT/SMT), as a paradigmatic quantum phase transition, has been widely studied over the past thirty years[11-14]. One striking phenomenon that contradicts the conventional wisdom during the SIT/SMT is the intermediate anomalous metallic state. Owing to Anderson localization, 2D metallic states are prohibited at zero temperature[15]. Consistent with this, there are only superconducting and insulating ground states and a direct transition between them in the framework of the dirty-boson model[16].

However, pioneering works on amorphous thin films, as well as recent experiments on crystalline 2D systems report the possible signature of intervening anomalous metallic state, characterized by the saturating resistance much lower than the normal state resistance[3, 4, 17-23]. Until now, the existence and the underlying mechanism of the anomalous metallic state are still under intensive and wide debate, partially due to its subtle nature and the possible influence of external noise in experiments[24]. Especially, such anomalous metallic states have not been reported in high quality 2D crystalline superconducting films grown by MBE. Thus, more convincing evidences on this intriguing phenomenon are highly desired, especially in 2D crystalline systems that host robust superconductivity.

Furthermore, the in-plane critical field of 2D superconductors is normally determined by the spin pair breaking effect, i.e. the Pauli limit. The broken inversion symmetry can induce strong SOC fields in the superconducting systems, which significantly enhances the superconducting upper critical field beyond the Pauli limit[25, 26]. For instance, Ising superconductivity with a large in-plane critical field was reported in several 2D crystalline superconductors with in-plane inversion symmetry breaking[5-7, 27-29]. The broken in-plane inversion symmetry can give rise to the Zeeman-type SOC, which polarizes the spins of electrons to the out-of-plane orientation and protects the superconductivity under large parallel magnetic field. The discovery of Ising superconductivity inspires further investigations on various 2D materials. However, the in-plane inversion symmetry is preserved in most 2D superconducting systems, where the aforementioned Ising superconductivity cannot exist.

Transition metal dichalcogenide $PdTe_2$ crystal has been demonstrated to be type-II Dirac semimetal[30] and exhibits a superconducting transition at 1.7 K[31], which makes it promising to explore novel physics (e.g. topological superconductivity). Reduced dimensionality, quantum fluctuation, as well as SOC effects in 2D PdTe$_2$ films may lead to unexpected quantum phenomena. Here we report systematic ultralow temperature transport study on macro-size ambient-stable superconductors with strong SOC, i.e. atomically-flat ultrathin crystalline PdTe$_2$ films on SrTiO$_3$(001) substrate prepared by MBE. The superconductivity in these PdTe$_2$ films with in-plane inversion symmetry survives a large parallel magnetic field more than 7 times of the Pauli limit, indicating an anomalous Zeeman-protected Ising superconductivity. Our electronic structure calculations and theoretical analysis unveil a new mechanism of Ising superconductivity without inversion symmetry breaking (named type-II Ising superconductivity) in 2D materials. Furthermore, by performing ultralow temperature transport measurement down to 15 mK, the sheet resistance of ultrathin PdTe$_2$ film drops and then saturates with decreasing temperature at finite perpendicular magnetic fields, showing the characteristic of anomalous metallic state. High quality filters are used in our measurements, which can exclude more than 99.999% (-50 dB) of the high frequency noise power, demonstrating that the observed anomalous metallic state is intrinsic. Our work reveals robust superconductivity with type-II Ising paring and provides solid evidence of anomalous metallic state in macroscale ambient-stable ultrathin crystalline films.

The ultrathin crystalline PdTe$_2$ films were grown on 0.05 wt% Nb-doped SrTiO$_3$(001) substrates via MBE technique (See Methods for details). Figure 1a shows a schematic of lattice structure of layered PdTe$_2$. It is composed of inversion-symmetric Te-Pd-Te layers that show AA stacking along the (0001) direction[32]. As confirmed by scanning tunneling microscopy (STM) measurements (a typical STM image of 4-monolayer (ML) PdTe$_2$ film is shown in Figure S1), the in-plane structure has a uniform hexagonal lattice with constant of 4.0 ± 0.1 Å. The 4-ML and 6-ML PdTe$_2$ films have been observed to be superconducting with respective critical transition temperature $T_c$ (defined as the temperature required to reach 50% of the normal state sheet resistance $R_n$) of 701.5 mK and 734.7 mK (Figure S3a). Figures 1b and 1c display the in-plane magnetoresistance of 4-ML and 6-ML PdTe$_2$ films up to 9 T. We define in-plane critical field $B_c^{\parallel}$ as the magnetic field corresponding to 50% of $R_n$ (dashed lines in Figures 1b and 1c) and estimate the Pauli limit by $B_P$ (in tesla) $= 1.84 T_c$ (in Kelvin)[33, 34], which is 1.29 T for 4-ML and 1.36 T for 6-ML PdTe$_2$ films. The temperature dependence of $B_c^{\parallel}$ normalized by $B_P$ shows a good agreement with the microscopic theory of Ising superconductivity, yielding effective out-of-plane Zeeman-type SOC $\widetilde{\beta_{SO}} = 1.09$ meV for 6-ML and 1.85 meV for 4-ML PdTe$_2$ films (see Supporting Information Part III for details). Here the effective Zeeman-type SOC $\widetilde{\beta_{SO}} = \beta_{SO}/[1 + \hbar/(2\pi k_B T_c \tau_0)]$, which takes the average Zeeman-type SOC $\beta_{SO}$ and the spin-independent scattering into account[28] ($\tau_0$ is the mean free time for spin-independent scattering). In theory, we also generalize the one-band Ising superconductivity to the two-band case, since PdTe$_2$ film is a two-band superconductor (see Supporting Information Part IV for details).

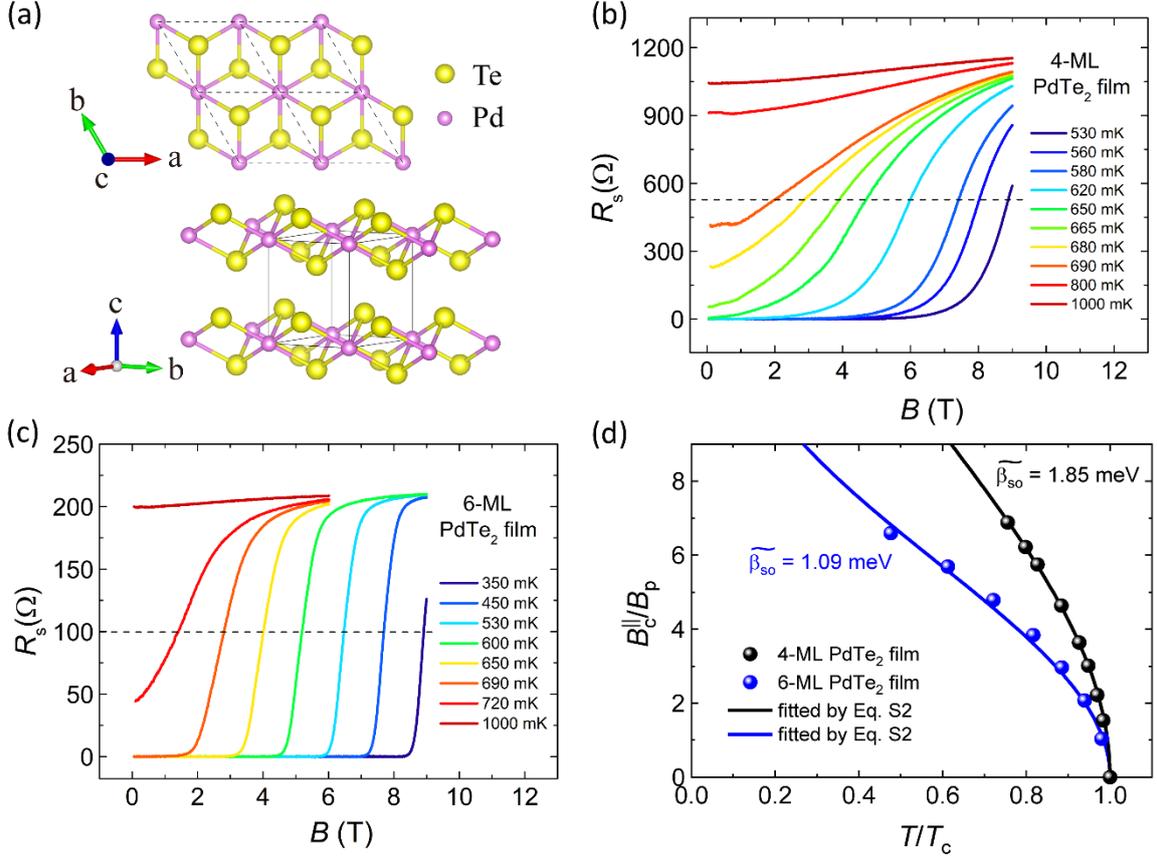

**Figure 1.** Large in-plane critical field of 4-ML and 6-ML PdTe$_2$ films. (a) The schematic of lattice structure of PdTe$_2$ with in-plane inversion symmetry. (b) The parallel magnetic field dependence of the sheet resistance in 4-ML PdTe$_2$ film at different temperatures ranging from 530 mK to 1000 mK. (c) The $R_s(B)$ curves of 6-ML PbTe$_2$ film from 350 mK to 1000 mK. (d) The in-plane critical field $B_c^{\parallel}$ of 4-ML (the black spheres) and 6-ML (the blue spheres) PbTe$_2$ films. Here, $B_c^{\parallel}$ is defined as the magnetic field required to reach the 50% of the normal state sheet resistance $R_n$ (the dashed lines in Figure 1b and 1c). The solid blue and black lines represent the fitting curves using the theoretical formula for Ising superconductor (Eq. S2).

Previously reported Ising superconductivity originates from the in-plane inversion symmetry breaking[5-7], which gives rise to pronounced Zeeman-type spin splitting in the bands around the K and K′ point of the Brillouin zone. Surprisingly, the PdTe$_2$ ultrathin films studied here are in-plane centrosymmetric (Figure 1a). Thus, the aforementioned Ising superconductivity mechanism for non-centrosymmetric systems[5-7] (named type-I Ising superconductivity) cannot explain our observations in PdTe$_2$ films. Instead, a new mechanism of Ising superconductivity (type-II Ising superconductivity) is proposed to explain our observations, which does not require in-plane inversion symmetry breaking. For four-fold degenerate $p_{xy}$ orbitals at the Γ point, the SOC introduces a so-called spin-orbital locking $H_{\text{SOC}} = \beta_{\text{SO}} \tau_z \sigma_z$, where $\beta_{\text{SO}}$ represents the Zeeman-type SOC strength and $\sigma_z, \tau_z = \pm 1$ label spin up/down and $p_x \pm i p_y$ orbitals, respectively. The Zeeman-type SOC splits the $p_{xy}$ orbitals into two sets of two-fold degenerate bands. As illustrated in Figure 2a, the $p_x + i p_y$ orbit is marked as the solid curves and $p_x - i p_y$ orbit is marked as the dashed curves. The spins of the electrons are shown by different colors (red for spin-up and blue for spin-down). The effective Zeeman field $\beta_{\text{SO}} \tau_z$ of SOC is time-reversal invariant, showing opposite signs for the two opposing orbitals. Importantly, its direction is out-of-plane as ensured by the $C_3$ rotational symmetry and its magnitude can be extraordinarily large due to the strong SOC. Electron spins are thus strongly polarized along the out-of-plane direction. Notably, although large spin splitting exists around the Γ point, the time-reversal symmetry is preserved since the spin-up electrons from $p_x + i p_y$ orbit and spin-down electrons from $p_x - i p_y$ orbit are degenerate, which form the Cooper pairs. Figure 2b displays the band structures around the Γ point of 6-ML PdTe$_2$ with and without SOC, which visualizes the SOC-induced splitting for $p_{xy}$ orbitals. When considering the SOC, the four-fold degenerate bands contributed by $p_{xy}$ orbitals (the black dashed curves denoted by the cyan dots) at the Γ point split into two sets of spin-degenerate bands (the blue solid curves with red dots above and below the cyan dots), consistent with schematic diagram of type-II Ising superconductivity in Figure 2a (See Figure S6 for details of the spin splitting in the bandstructure). First-principles calculations reveal that spin splitting induced by in-plane magnetic field is considerably suppressed by Zeeman-type SOC, and the suppression is prominent near Γ and gets weaker with increasing momentum (Figure 2c), indicating that the protection of Zeeman-type SOC mainly comes from the bands close to the Γ point. Despite of several bands across the Fermi surface, a mean-field $\widetilde{\beta_{SO}}$ can be used as an approximation to consider the overall effects for a given Fermi energy. Thus, the superconducting states are protected against large in-plane magnetic field by the out-of-plane Zeeman-type SOC field, displaying the in-plane critical field far beyond the Pauli limit.

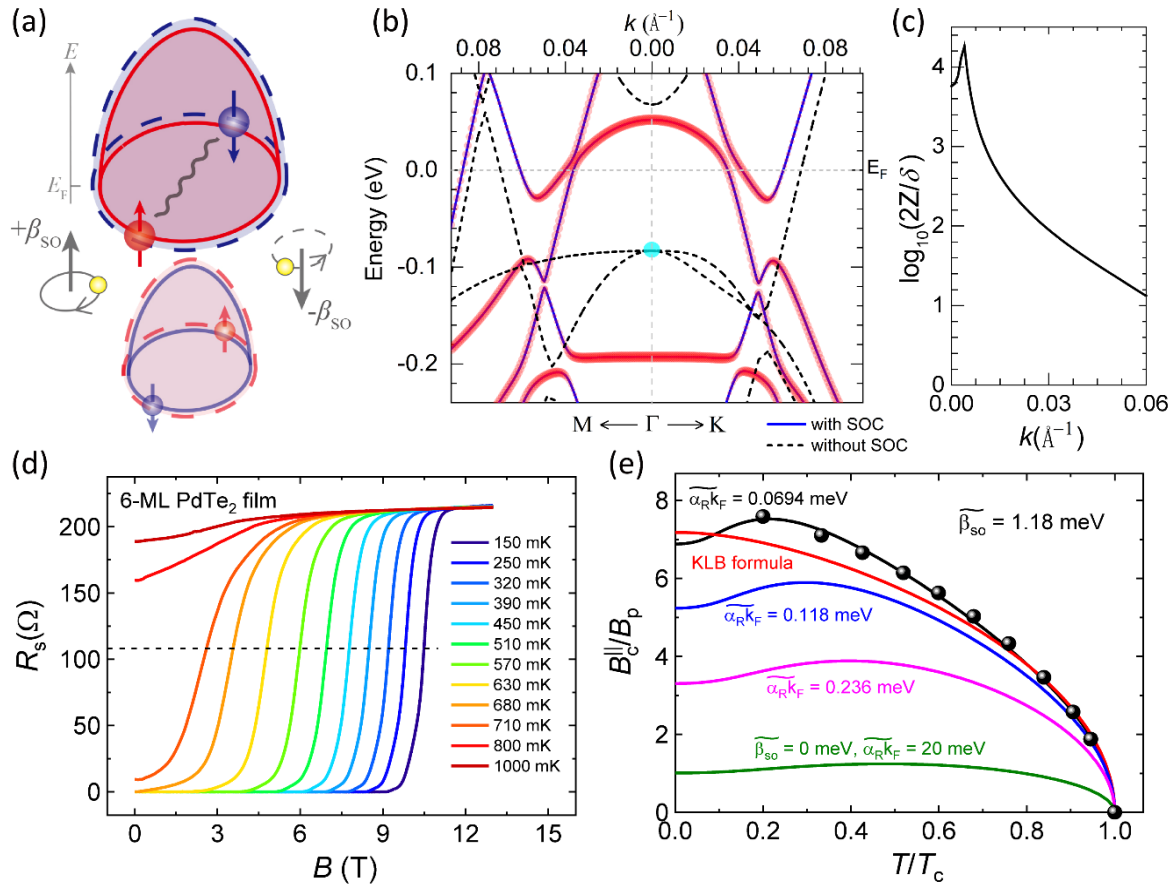

**Figure 2.** Type-II Ising superconductivity in ultrathin PdTe$_2$ films. (a) Schematic diagram of type-II Ising superconductivity mechanism. For multiple degenerate orbitals near a high-symmetry point (Γ point), the SOC results in a spin-orbital locking and generates opposite Zeeman-like fields ($\pm\beta_{SO}$) for the two opposing orbitals (dashed and solid lines), which strongly polarizes spins (red and blue) along the out-of-plane direction. Type-II Ising superconducting paring formed between opposite spins, momenta and orbitals is thus protected against in-plane magnetic fields, displaying unusually large upper critical fields. (b) Band structures around the Γ point of 6-ML PdTe$_2$ with and without SOC. Bands contributed by $p_{xy}$ orbitals are four-fold degenerate at Γ without SOC (the black dashed curves denoted by cyan dots) and become split into two sets of spin-degenerate bands with SOC (the blue solid curves with red dots above and below the cyan dots) (See Figure S6 for details of the spin splitting in the bandstructure). The bandstruture around the Γ point is consistent with the schematic of type-II Ising superconductivity. (c) The calculated ratio between Zeeman splitting without SOC (2Z) and with SOC (δ) as a function of momentum (k) referenced to Γ for a near-Fermi-level band of 6-ML PdTe$_2$ contributed by $p_{xy}$ orbitals under an in-plane magnetic field of two Tesla. (d) The parallel magnetic field dependence of the sheet resistance in 6-ML PdTe$_2$ film at different temperatures from 150 mK to 1000 mK. (e) Theoretical fitting of $B_c/B_p$ as a function of $T/T_c$ using a fixed effective Zeeman-type SOC and increasing effective Rashba-type SOC. The black curve presents the best fit of the experimental data with effective Zeeman-type SOC $\widetilde{\beta_{SO}} = 1.18$ meV and effective Rashba-type SOC $\widetilde{\alpha_R k_F} = 0.0694$ meV. The olive line indicates a special case with only effective Rashba-type SOI ($\widetilde{\beta_{SO}} = 0, \widetilde{\alpha_R k_F} = 20$ meV), which is slightly above the Pauli limit. The red curve represents the theoretical fitting using the KLB formula (spin-orbit scattering mechanism), which is not consistent with our experimental

data at low temperatures.

To further investigate the large in-plane critical field at lower temperatures, we performed an ultralow temperature measurement on the same 6-ML PdTe$_2$ film up to 13 T. This measurement was carried out around 20 months after the growth of the sample and surprisingly the superconducting properties (Figures 2d and 2e) remain almost the same as the previous data in Figures 1c and 1d (one month after growth), demonstrating robust superconductivity in ultrathin PdTe$_2$ films without any capping layer. In Figure 2e, the normalized in-plane critical field $B_c^{\parallel}/B_P$ as a function of $T/T_c$ can be well fitted by the microscopic formula (Eq. S3) with two fitting parameters effective Zeeman-type and Rashba-type SOC (see Supporting Information Part III for details). The effective Zeeman-type SOC $\widetilde{\beta_{SO}}$ (1.18 meV) is much larger than the effective Rashba-type SOC $\widetilde{\alpha_R k_F}$ (0.0694 meV) in 6-ML PdTe$_2$ film. In the Ising superconducting systems, the spin orientation is polarized to the out-of-plane direction by Zeeman-type SOC. However, Rashba-type SOC weakens the spin orientation by tilting the electron spin to the in-plane direction, which can be more easily affected by the parallel magnetic field. To demonstrate this, we plot a serial of theoretical curves with a fixed $\widetilde{\beta_{SO}}$ of 1.18 meV and increasing $\widetilde{\alpha_R k_F}$ from 0.0694 to 0.236 meV in Figure 2e. The theoretical curves with larger effective Rashba-type SOC exhibit smaller in-plane critical fields and significantly deviate from our observation. A special case with only large effective Rashba-type SOC ($\widetilde{\alpha_R k_F}$ =20 meV) is also considered, where the in-plane critical field is slightly above the Pauli limit, indicating that the Rashba-type SOC alone cannot explain the enhanced $B_c^{\parallel}$ of ultrathin PdTe$_2$ film. Then we discuss other possible origin of large in-plane critical field for 2D superconducting systems. The Pauli limit can be surpassed by spin-orbit scattering mechanism[35], since it randomizes the electron spins and weakens the spin pair breaking effect of external in-plane magnetic field. It is noteworthy to mention that the spin-orbit scattering mechanism, described by the Klemm-Luther-Beasley (KLB) formula[35] (the solid red curve in Figure 2e) deviates from the enhanced $B_c^{\parallel}$ of 6-ML PdTe$_2$ films in the low temperature regime when $T < 0.6 T_c$. Although spin-orbit scattering may contribute to the enhanced in-plane critical field in 2D centrosymmetric superconductors[29], our observation is qualitatively different from this mechanism.

Figure 3a presents the Arrhenius plot of the sheet resistance $R_s(T)$ curves in the temperature from 15 mK to 1 K for 4-ML PdTe$_2$ film at different perpendicular magnetic fields. In the relatively high temperature regime, $\log R_s$ scales linearly with $T^{-1}$, indicative of an activated behavior, which is usually attributed to the thermally assisted motion of unpaired vortex, namely the thermal creep[36]. As shown in Figure 3b, the activation energy $U(B)$ at various magnetic fields are determined by the slope of the linear portion (the dashed black line) in Figure 3a and can be well fitted by the formula[36] $U(B) = U_0 \ln(B_0/B)$ with $U_0 = 2.64$ K and $B_0 = 0.613$ T for the 4-ML PdTe$_2$ film. When the magnetic field is approaching 0.613 T, the activation energy $U$ drops around zero indicating weak pinning of vortices.

In the ultralow temperature regime, when the field is above 0.491 T, the sheet resistance significantly deviates from the activated behavior below $T_{AM}$ (marked as the arrows in Figure 3a) and then saturates to a finite value with decreasing temperature down to 15 mK, which is the hallmark of anomalous metallic state. Here, well-filtered electrical leads (Figures 3c and 3d) are used in our measurement to investigate the anomalous metallic state with linear IV curves (the inset of Figure 3b). The attenuation of the filters can be expressed as $10\lg(P_\text{o}/P_\text{i})$, where $P_\text{i}$ and $P_\text{o}$ represent the power of the high frequency noise before and after the filters. In Figure 3d, the thermocoax filter and resistor-capacitor (RC) filter used in our measurement provide a good performance better than -50 dB, indicating that more than 99.999% of high frequency noise power is reduced. The performance of the filters becomes even better at higher frequency and reaches -100 dB at 1.1 GHz. Therefore, the extrinsic origin of the resistance saturation due to the high frequency noise, reported in recent experiments on 2D NbSe$_2$ and InO$_\text{x}$ superconductors[24], can be safely excluded, which makes our detection of anomalous metallic state undoubted. Moreover, we measured the magnetoresistance of the same sample at ultralow temperatures, which is consistent with the theoretical model of quantum tunneling of vortices (quantum creep) (Figure 3e). In the limit of strong dissipation, the sheet resistance of this model can be expressed as follow[4, 37]:

$$R_s \sim \frac{\hbar}{4e^2} \frac{\kappa}{1-\kappa} \quad [1]$$

$$\kappa \sim \exp[C \frac{\hbar}{e^2} \frac{1}{R_\text{n}} \left(\frac{B - B_\text{c2}^\perp}{B}\right)]$$

where C is a dimensionless constant of order unity, $R_\text{n}$ is the sheet resistance of the normal state (874.5 Ω for 4-ML PdTe$_2$ film) and $B_\text{c2}^\perp$ is the perpendicular upper critical field. As shown in Figure 3e, the magnetoresistance at low temperatures (below 100 mK) and relatively low magnetic fields is well fitted by Eq. 1, indicating that the quantum creep plays a key role in this regime as a possible origin of anomalous metallic state (See Table S3 for the fitting parameters). While at larger magnetic fields above 0.63 T, the sheet resistance at 20 mK exhibits a linear magnetic field dependence, showing the characteristic of pinning-free vortex motion (Figure 3f).

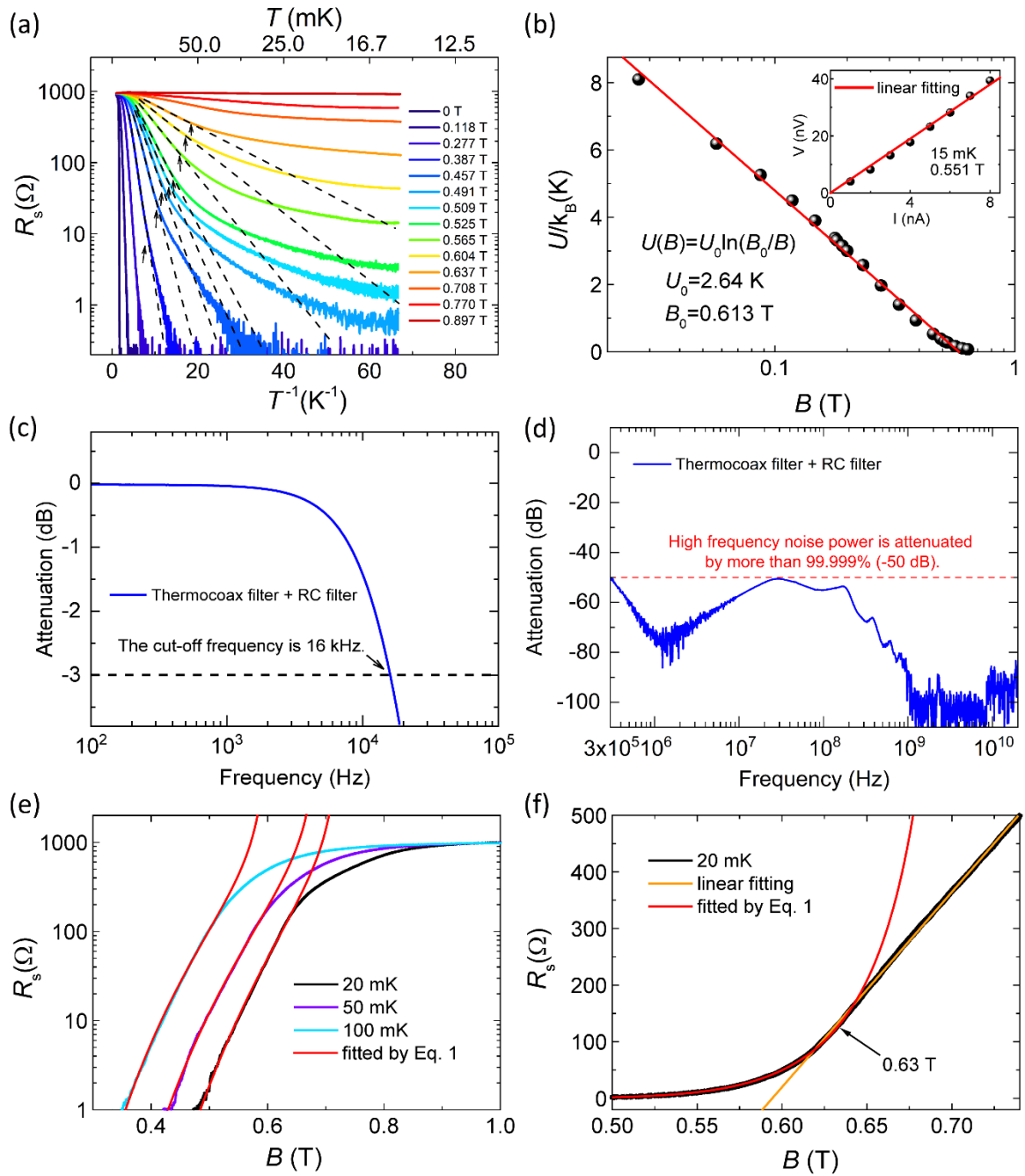

**Figure 3.** Evidence for thermally assisted vortex motion and anomalous metallic state in 4-ML PdTe$_2$ film. (a) Arrhenius plot of the sheet resistance at different perpendicular magnetic fields

from 0 T to 0.897 T. The dashed black lines show the activated behavior in the relatively high temperature regime. (b) The magnetic field dependence of activation energy $U$, obtained from the slope of the linear portion in Figure 3a. The inset: the linear IV curve at 20 mK and 0.551 T. (c) The performance of high-quality filters (Thermocoax filter and resistor-capacitor (RC) filter). The cut-off frequency is around 16 kHz. (d) Above 300 kHz, more than 99.999% (-50 dB) of the high frequency noise power is filtered. Above 1.1 GHz, the filters provide an even better performance and reach the analyzer's noise floor (around -100 dB). (e) The $R_s(B)$ curves of 4-ML PbTe$_2$ film at 20, 50 and 100 mK, which can be well fitted by Eq. 1 in the relatively low magnetic field regime (the solid red lines). (f) The $R_s(B)$ curve exhibits a linear magnetic field dependence above 0.63 T.

In summary, we detect large in-plane critical field more than 7 times of the Pauli limit in the ultrathin crystalline PdTe$_2$ films, which can be well explained by a new kind of Ising superconductivity (named type-II Ising superconductivity) in the 2D centrosymmetric systems. Under perpendicular magnetic field, the sheet resistance of PdTe$_2$ film drops and then saturates with decreasing temperatures, which is the key signature of the anomalous metallic state. High quality filters and linear IV characteristics demonstrate that the observed anomalous metallic state is intrinsic. Importantly, this macroscale transition metal dichalcogenide PdTe$_2$ films down to a few monolayers show ambient-stable superconductivity with interesting quantum phases, which not only offer an ideal platform for further investigations on novel quantum phenomena but also exhibit great potentials in electronic and spintronic applications (e.g. integrated electronic devices and spintronics with persistent spin textures[38]).

## ASSOCIATED CONTENT

**Supporting Information**

The following file is available free of charge.

Methods, theoretical fitting for perpendicular upper critical field, theoretical formula for one-band Ising superconductivity, theoretical formula for two band type-II Ising superconductivity, supporting Figures and Tables.

## AUTHOR INFORMATION

**Corresponding Author**

*E-mail: jianwangphysics@pku.edu.cn (J.W.).

*E-mail: liliwang@mail.tsinghua.edu.cn (L.W.).

*E-mail: xilin@pku.edu.cn (X.L.).

**Author Contributions**

J.W. conceived the research. J.W., L.W. and X.L. instructed the experiments. C.L., C.D., L.W. and Q.-K.X. grew the samples. Yi Liu, J.S., Yanzhao Liu, K.G., Y.T., X.L. and J.W. performed the transport measurements and analyzed the experimental results. H.L. and Y.X. developed the theoretical model. Y.X., C.W. and Z.Z. carried out first-principles calculations. Yi Liu, J.S., Yanzhao Liu, Z.Z., L.W. and J.W. wrote the paper with suggestions from all other authors.

‡Yi Liu, Y. X., J. S. and C. L. contributed equally to this work.

**Notes**

The authors declare no competing financial interest.

ACKNOWLEDGMENT

We thank X. C. Xie, Wenhui Duan for helpful discussions and Pengjie Wang, Yifang Xu, Qingzheng Qiu, Yaochen Li for the help in ultralow temperature transport measurement. This work was financially supported by the National Key Research and Development Program of China (Grant No. 2018YFA0305600, No. 2017YFA0303300, No. 2015CB921000, No. 2015CB921100 and No. 2018YFA0307100), the National Natural Science Foundation of China (Grant No. 11888101, No.11774008, No. 11774193, No. 11790311, No. 51788104, No. 11874035 and No. 11674188), the Strategic Priority Research Program of Chinese Academy of Sciences (Grant No. XDB28000000), Beijing Natural Science Foundation (Z180010), the Beijing Advanced Innovation Center for Future Chip (ICFC) and China Postdoctoral Science Foundation (Grant No. 2019M650290).

Supporting Information for

# Type-II Ising superconductivity and anomalous metallic state in macro-size ambient-stable ultrathin crystalline films

*Yi Liu[†,Δ,‡], Yong Xu[Δ,§,||,‡], Jian Sun[†,Δ,‡], Chong Liu[§,‡], Yanzhao Liu[†,Δ], Chong Wang[Δ,§,⊥], Zetao Zhang[Δ,§], Kaiyuan Gu[†,Δ], Yue Tang[†,Δ], Cui Ding[§], Haiwen Liu[#], Hong Yao[Δ,§,⊥], Xi Lin[†,Δ,//,¶,\*], Lili Wang[Δ,§,\*], Qi-Kun Xue[Δ,§,//] & Jian Wang[†,Δ,§,//,¶,\*]*

[†]International Center for Quantum Materials, School of Physics, Peking University, Beijing 100871, China.

[Δ]Collaborative Innovation Center of Quantum Matter, Beijing 100871, China.

[§]State Key Laboratory of Low-Dimensional Quantum Physics, Department of Physics, Tsinghua University, Beijing 100084, China.

[||]RIKEN Center for Emergent Matter Science, Wako, Saitama 351-0198, Japan

[⊥]Institute for Advanced Study, Tsinghua University, Beijing 100084, China.

[#]Center for Advanced Quantum Studies, Department of Physics, Beijing Normal University, Beijing 100875, China.

[//]Beijing Academy of Quantum Information Sciences, Beijing 100193, China.

[¶]CAS Center for Excellence in Topological Quantum Computation, University of Chinese Academy of Sciences, Beijing 100190, China.

[‡]These authors contributed equally to this work.

*Corresponding Author. E-mail: jianwangphysics@pku.edu.cn (J.W.);

liliwang@mail.tsinghua.edu.cn (L.W.); xilin@pku.edu.cn (X.L.).

Contents:

I. Methods

II. Theoretical fitting for perpendicular upper critical field

III. Theoretical formula for one-band Ising superconductivity

IV. Theoretical formula for two band type-II Ising superconductivity

V. Figures and Tables

I. Methods

**Sample synthesis.** The 0.05 wt.% Nb-doped $SrTiO_3$(001) was chosen as substrates, and the $TiO_2$-terminated surface was obtained after being heated to 1100 °C in ultrahigh vacuum. The $PdTe_2$ films were grown by co-evaporating Pd (99.995%) and Te (99.9999%) from Knudsen cells with a flux ratio of ~1:2 on the $SrTiO_3$ substrates kept at 120 °C. The growth rate was approximately 0.07 ML per minute. Scanning tunneling microscopy (STM) images were acquired at 4.8 K in a constant current mode, with the bias voltage ($V_s$) applied to the sample.

**Ultralow temperature transport measurements.** The experiments for anomalous metallic state were carried out in a dilution refrigerator (MNK 126-450; Leiden Cryogenics BV) with a base refrigerator temperature below 6 mK. The thermometry was based on a commercial thermometer and checked by a cerium magnesium nitrate susceptibility thermometer. At temperature ≥ 15 mK, the refrigerator temperature and the electron temperature were measured to be equal with use of the temperature dependence of fractional quantum Hall states' longitudinal resistances.

The experiments investigating the type-II Ising superconductivity were performed in another dilution refrigerator (CF-CS81-600; Leiden Cryogenics BV) and a commercial Physical Property Measurement System (Quantum Design, PPMS-16) with the dilution refrigerator option. The parallel magnetic field condition in CF-CS81-600 system was realized by a piezo-driven sample rotation system (ANR101; Attocube AG) with a readout resolution of ±0.02° and stability of ±0.05°. We checked the thermometry and the electron temperature in the same way with that in MNK 126-450 system, and the electron temperature of the samples is consistent with the refrigerator temperature when above 25 mK.

In the above-mentioned measurements in MNK 126-450 and CF-CS81-600 systems, standard lock-in techniques were used and the alternating current excitations were selected according to the criteria of linear I-V curves. The transport properties of the macroscopic $PdTe_2$ films were measured with the standard four-probe method by using indium as the electrodes. Two narrow strips of current electrodes (I+ and I-) were pressed along the width of the sample to provide a homogeneous current in the films and the other two electrodes served as the voltage probes (V+ and V-). To filter out external radiation and obtain ultralow electron temperature environment, filters have been used. In MNK 126-450 system, Thermocoax filters (installed from room temperature to sample plate) and home-made low temperature resistor-capacitor (RC) filters (installed at sample plate) are connected in series with each lead of samples for the detection of anomalous metallic state. The performance of the filters was tested via a vector network analyzer (E5071C ENA; Agilent Technologies) at room temperature. Thermocoax and RC filters provide a good performance better than -50 dB within the frequency measurement range of the analyzer and reaches the analyzer's noise floor for frequency above 1.1 GHz (Figure 3d). The cut-off frequency of thermocoax and RC filters is 16 kHz measured via a lock-in amplifier (SR830; Stanford Research Systems). While in CF-CS81-600 system, home-made room temperature RC filters and home-made silver-epoxy filters (installed at the sample plate) are connected in series with each lead of samples for the investigation of type-II Ising superconductivity. The silver-epoxy and RC filters provide a good performance better than -45 dB within the frequency measurement range of the analyzer and reach the analyzer's noise floor (around -100 dB) for frequency above 200 MHz. The cut-off frequency of silver-epoxy and RC filters is 6 kHz.

**First-principles calculations.** First-principles calculations based on density functional theory (DFT) were performed by the *OpenMX* package [1,2] using norm-conserving pseudo-potentials and the Perdew-Burke-Ernzerhof exchange-correlation functional [3]. Bloch wavefunctions were expanded by numerical pseudo-atomic orbitals with optimized radial part *s*2*p*2*d*2*f*1 (*s*3*p*3*d*2*f*1) and a cutoff radius of 7.0 Bohr for Pd (Te) atoms. SOC was included with fully relativistic scheme and *j*-dependent pseudo potentials. Brillouin zone was discretized by a 14×14×1 grid. Convergence criterion of self-consistent calculations was selected to be $10^{-9}$ Hartree. The geometry structures studied by OpenMX were optimized by Vienna *ab initio* simulation package (VASP) [4], using a slab model with a vacuum layer over 15Å and DFT-D3 method [5] for van der Waals corrections. The Kohn-Sham Hamiltonian was read from the output of OpenMX and a Zeeman term of in-plane magnetic field was added to calculate the energy splitting of Kramer pairs. At a specific distance away from Γ point, 10 k-points were uniformly selected around Γ to calculate the Zeeman splitting and the averaged splitting was chosen as the Zeeman splitting at a specific distance.

## II. Theoretical fitting for perpendicular upper critical field.

The theoretical formula for the perpendicular upper critical field of two band superconductors can be expressed as [6]

$$\frac{2\omega}{\lambda_0} f(t,h)f(t,\eta h) + \left(1 + \frac{\lambda_-}{\lambda_0}\right) f(t,\eta h) + \left(1 - \frac{\lambda_-}{\lambda_0}\right) f(t,h) = 0 \qquad [S1]$$

with $f(t,h) = \ln t + \left[\psi\left(\frac{1}{2} + h\right) - \psi(h)\right]$, $t = \frac{T}{T_c}$, $\eta = \frac{D_2}{D_1}$, $h = \frac{B_{c2}^\perp D_1}{2\phi_0 T}$,

and $\omega = \lambda_{11}\lambda_{22} - \lambda_{12}\lambda_{21}$, $\lambda_- = \lambda_{11} - \lambda_{22}$, $\lambda_0 = \sqrt{\lambda_-^2 + 4\lambda_{12}\lambda_{21}}$, $\lambda_{12} = \lambda_{21}$

where $\psi(x)$ is the digamma function. $D_1$ and $D_2$ represent intraband diffusivities of bands 1 and 2. $\phi_0$ denotes the magnetic flux quantum. Parameters $\lambda_{11}$ and $\lambda_{22}$ represent intraband couplings in band 1 and 2. $\lambda_{12}$ and $\lambda_{21}$ denote interband couplings between the two bands. As shown in Figure S3, the perpendicular upper critical field of 4-ML and 6-ML PdTe$_2$ films can be well fitted by Eq. S1 and the fitting parameters are shown in Table S1.

## III. Theoretical formula for one-band Ising superconductivity.

The microscopic model for the one-band Ising superconductivity to determine the in-plane critical field $B_c^\parallel(T)$ in terms of effective Zeeman-type SOC $\widetilde{\beta_{SO}}$ can be expressed as [7]:

$$\ln\left(\frac{T}{T_c}\right) + \frac{\mu_B^2 B^2}{\widetilde{\beta_{SO}}^2 + \mu_B^2 B^2} Re\left[\psi\left(\frac{1}{2} + \frac{i\sqrt{\widetilde{\beta_{SO}}^2 + \mu_B^2 B^2}}{2\pi k_B T}\right) - \psi\left(\frac{1}{2}\right)\right] = 0 \quad \text{[S2]}$$

where $\psi(x)$ is the digamma function, $\widetilde{\beta_{SO}} = \beta_{SO}/(1 + \frac{\hbar}{2\pi k_B T_c \tau_0})$ is the effective Zeeman-type SOC considering spin-independent scattering and $\tau_0$ is the mean free time of spin-independent scattering.

When considering the Rashba-type SOC in the one-band Ising superconductor [7], $B_c^\parallel$ as a function of $\frac{T}{T_c}$ is determined by both the effective Rashba-type SOC $\widetilde{\alpha_R k_F}$ and Zeeman-type SOC $\widetilde{\beta_{SO}}$:

$$ln\left(\frac{T}{T_c}\right) + \frac{1}{2}\left[1 - \frac{2(\widetilde{\alpha_R k_F})^2 + \widetilde{\beta_{SO}}^2 - \mu_B^2 B^2}{\rho_+^2 - \rho_-^2}\right] Re\left[\psi\left(\frac{1}{2} + \frac{i\rho_+}{2\pi k_B T}\right) - \psi\left(\frac{1}{2}\right)\right] + \frac{1}{2}\left[1 + \frac{2(\widetilde{\alpha_R k_F})^2 + \widetilde{\beta_{SO}}^2 - \mu_B^2 B^2}{\rho_+^2 - \rho_-^2}\right] Re\left[\psi\left(\frac{1}{2} + \frac{i\rho_-}{2\pi k_B T}\right) - \psi\left(\frac{1}{2}\right)\right] = 0 \quad \text{[S3]}$$

Where $\psi(x)$ is the digamma function, and the arguments are defined as

$$2\rho_\pm \equiv \sqrt{(\mu_B B + \widetilde{\alpha_R k_F})^2 + (\widetilde{\alpha_R k_F})^2 + \widetilde{\beta_{SO}}^2} \pm \sqrt{(\mu_B B - \widetilde{\alpha_R k_F})^2 + (\widetilde{\alpha_R k_F})^2 + \widetilde{\beta_{SO}}^2}$$

with effective Zeeman-type SOC $\widetilde{\beta_{SO}} = \beta_{SO}/(1 + \frac{\hbar}{2\pi k_B T_c \tau_0})$ and effective Rashba-type SOC $\widetilde{\alpha_R k_F} = \alpha_R k_F/\sqrt{2}(1 + \frac{\hbar}{2\pi k_B T_c \tau_0})$. $\tau_0$ is the mean free time of spin-independent scattering.

## IV. Theoretical formula for two band type-II Ising superconductivity.

Based on the perpendicular critical field analysis, we find the system behaves as a two band superconductor. Thus, we assume the superconductivity originates from the different energy band with two different Fermi wave-vector, and we mainly consider the intra-Fermi-surface scattering with two separate mean free time, and neglect the inter-Fermi-surface scattering [6]. The one-band Zeeman-protected Ising superconductivity can be obtained by the Gor'kov Green's function method [8, 7]. We can phenomenologically generalize the single band Zeeman-protected Ising superconductivity to the two-band case, and join the single band results together in light of the quasi-classical two-band Usadel equations [6]. Then, the two-band model for the in-plane critical field reads:

$$\frac{2\omega}{\lambda_0} F(\widetilde{\beta_{SO1}}, T, B) F(\widetilde{\beta_{SO2}}, T, B) + \left(1 + \frac{\lambda_-}{\lambda_0}\right) F(\widetilde{\beta_{SO1}}, T, B) + \left(1 - \frac{\lambda_-}{\lambda_0}\right) F(\widetilde{\beta_{SO2}}, T, B) = 0 \qquad [S4]$$

with

$$F(\widetilde{\beta_{SOr}}, T, B) = \ln\frac{T}{T_c} + \frac{\mu_B^2 B^2}{\widetilde{\beta_{SOr}}^2 + \mu_B^2 B^2} \left[ Re\psi\left(\frac{1}{2} + \frac{i\sqrt{\widetilde{\beta_{SOr}}^2 + \mu_B^2 B^2}}{2\pi k_B T}\right) - \psi\left(\frac{1}{2}\right) \right], \qquad (r = 1,2)$$

and $\omega = \lambda_{11}\lambda_{22} - \lambda_{12}\lambda_{21}$, $\lambda_- = \lambda_{11} - \lambda_{22}$, $\lambda_0 = \sqrt{\lambda_-^2 + 4\lambda_{12}\lambda_{21}}$, $\lambda_{12} = \lambda_{21}$,

where $\psi(x)$ is the digamma function. Parameters $\lambda_{11}$ and $\lambda_{22}$ represent intraband coupling in band 1 and 2. $\lambda_{12}$ and $\lambda_{21}$ denote interband interaction between the two bands. In theory, $\lambda_{11}$, $\lambda_{12}$, $\lambda_{21}$ and $\lambda_{22}$ are intrinsic parameters independent of the external magnetic field direction. To minimize the amount of fitting parameters, we obtained $\lambda_{11}$, $\lambda_{12}$, $\lambda_{21}$ and $\lambda_{22}$ from the theoretical fitting of the perpendicular upper critical field $B_{c2}^\perp$ of the PdTe$_2$ films, which also exhibits the characteristic of two band superconductivity (Figure S3 and Table S1). The parameters $\widetilde{\beta_{SO1}}$ and $\widetilde{\beta_{SO2}}$ are effective Zeeman-type SOC strength for band 1 and band 2, which give rise to the large in-plane critical field. As shown in Figure S5, the in-plane critical field $B_c^\parallel$ of 4-ML and 6-ML PdTe$_2$ films can be well fitted by Eq. S4, yielding the effective Zeeman-type SOC $\widetilde{\beta_{SO1}} = 1.836$ meV (band 1), $\widetilde{\beta_{SO2}} = 1.886$ meV (band 2) for 4-ML films, and $\widetilde{\beta_{SO1}} = 0.5077$ meV (band 1), $\widetilde{\beta_{SO2}} = 1.257$ meV (band 2) for 6-ML films, respectively (See Table S2 for all parameters). We also note that if the two values of Zeeman-type SOC parameters have little difference or the two values have very large difference, the Eq. S4 can go back to the formula of the one-band Zeeman-protected Ising superconductivity (Eq. S2).

## V. Figures and Tables

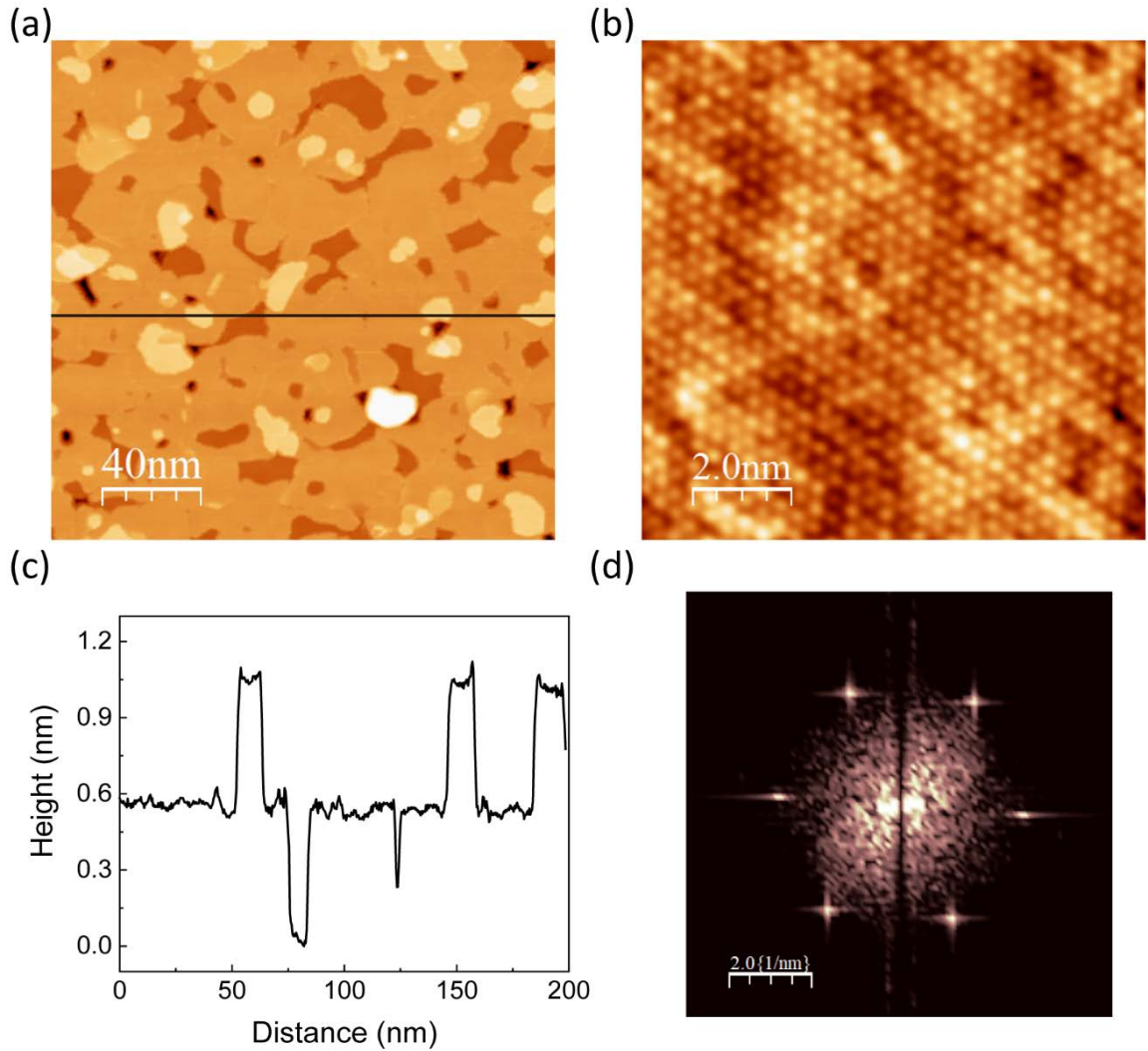

Figure S1. STM characterization of PdTe$_2$ films. (a) A typical STM topographic image of PdTe$_2$ film with a nominal coverage of 4 ML (200 nm × 200 nm, sample bias $V_s$ = 5 V, tunneling current $I_t$ = 100 pA) showing atomically flat crystalline films. (b) Atomically resolved STM image (10 nm × 10 nm, $V_s$ = 30 mV, $I_t$ = 1 nA) showing periodic hexagonal lattices with in-plane lattice constant of 4.0 ± 0.1 Å. (c) The line cut topographic profiles of Figure S1a. The small islands and holes on the sample surface are 1-ML thick. (d) FFT image of Figure S1b.

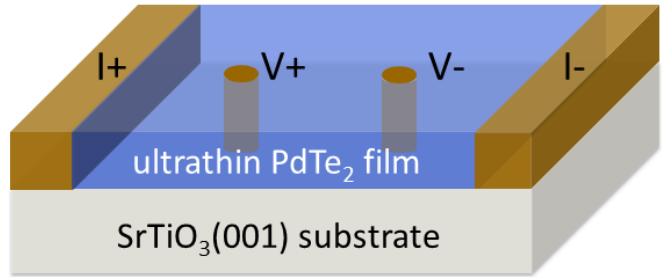

Figure S2. The schematic of the standard four-electrode transport measurement on macroscopic PdTe$_2$ films. Two narrow strips of indium were pressed as the current electrodes (I+ and I-) along the width of the samples to provide a homogeneous current in the films and the other two indium electrodes served as the voltage probes (V+ and V-).

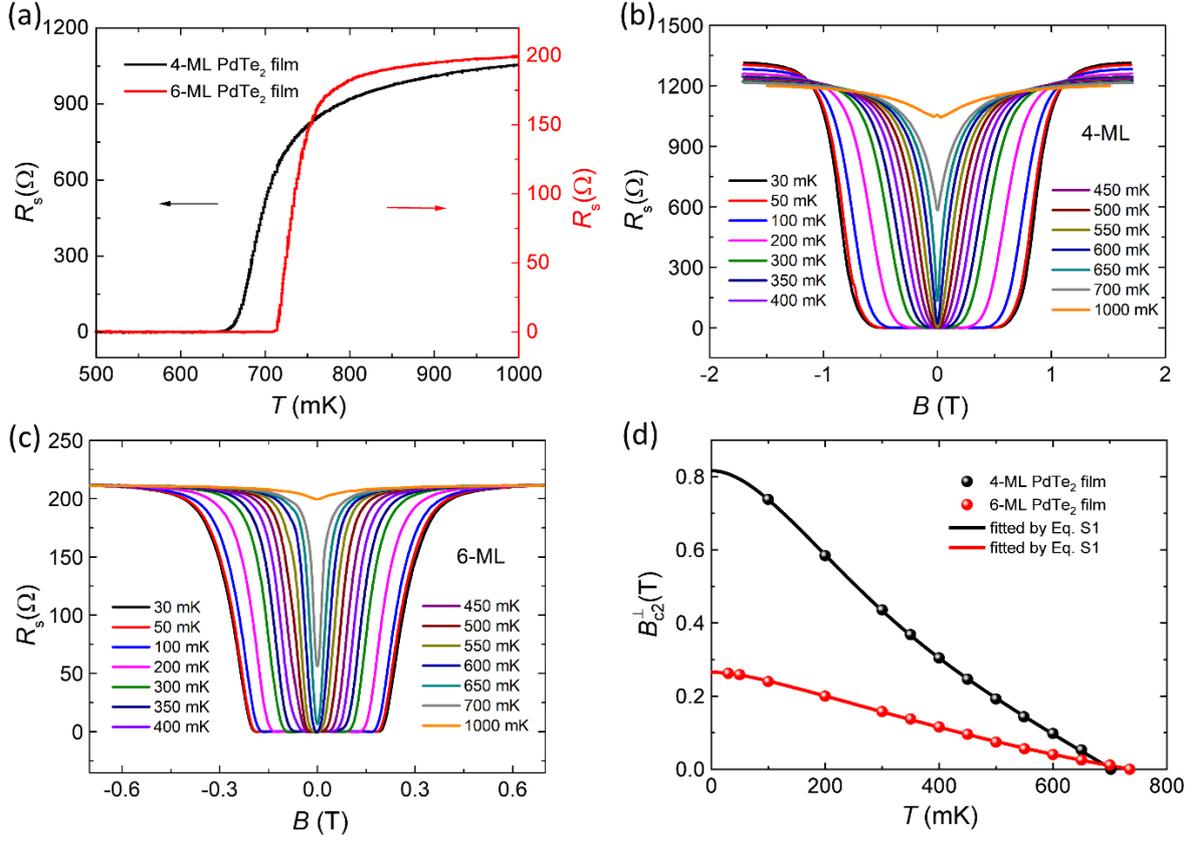

Figure S3. Superconductivity in ultrathin PdTe$_2$ films at perpendicular magnetic fields. (a) The temperature dependence of sheet resistance at 0 T, showing critical transition temperature $T_c$ of 701.5 mK for 4-ML and 734.7 mK for 6-ML PdTe$_2$ films. (b) (c) The perpendicular magnetic field dependence of the sheet resistance in 4-ML (b) and 6-ML (c) PdTe$_2$ films at different temperatures ranging from 30 mK to 1000 mK. (d) The upper critical field $B_{c2}^{\perp}$ of 4-ML (the black spheres) and 6-ML (the red spheres) PbTe$_2$ films. Here, $B_{c2}^{\perp}$ is defined as the magnetic field required to reach the 50% of the normal state sheet resistance $R_n$. The solid black and red lines represent the fitting curves using Eq. S1.

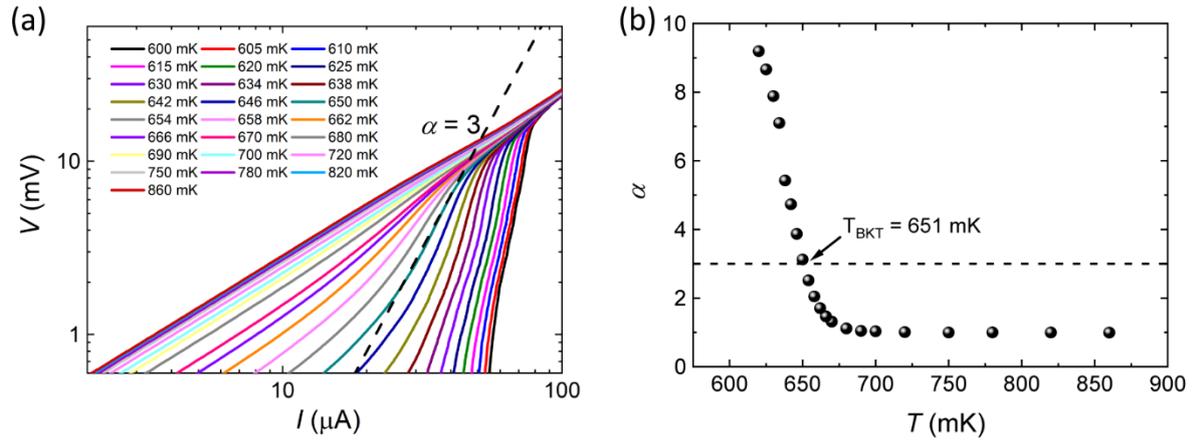

Figure S4. Berezinski-Kosterlitz-Thouless (BKT)-like transition in ultrathin PdTe$_2$ film. (a) *V*(*I*) curves at different temperatures from 600 mK to 860 mK for 4.5-ML PdTe$_2$ film, showing BKT-like transition. The black dashed line represent $V \propto I^3$ curve. (b) The temperature dependence of the exponent $\alpha$ extracted from Figure S4a, revealing the BKT transition temperature $T_{\text{BKT}} = 651$ mK.

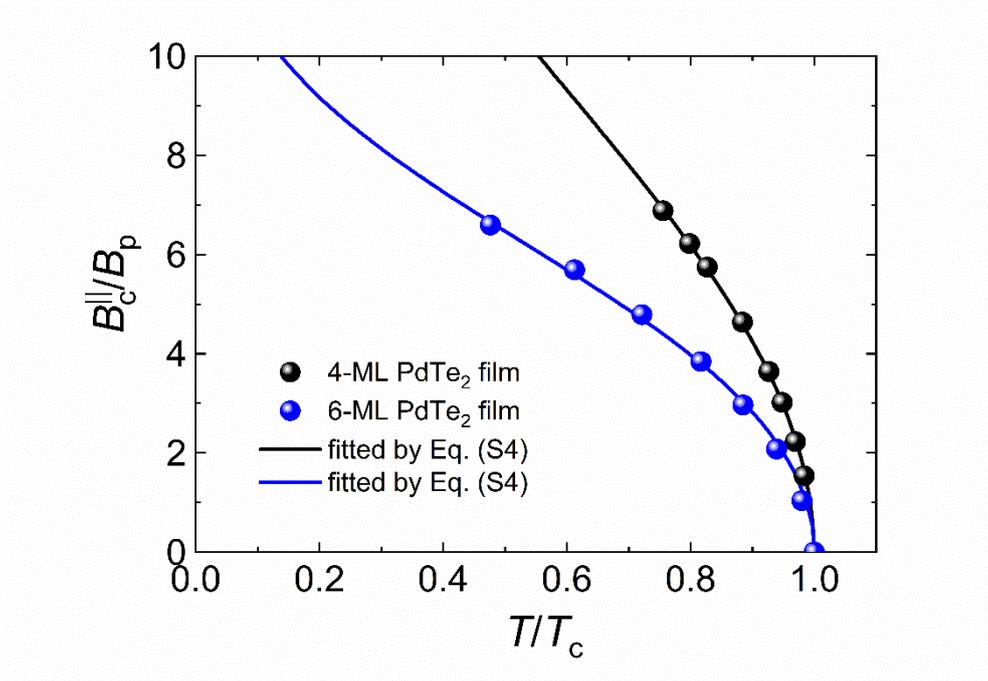

Figure S5. The in-plane critical field $B_c^{\parallel}$ of 4-ML (the black spheres) and 6-ML (the blue spheres) PbTe$_2$ film. The solid blue and black lines represent the fitting curves using the theoretical formula for two-band type-II Ising superconductor (Eq. S4).

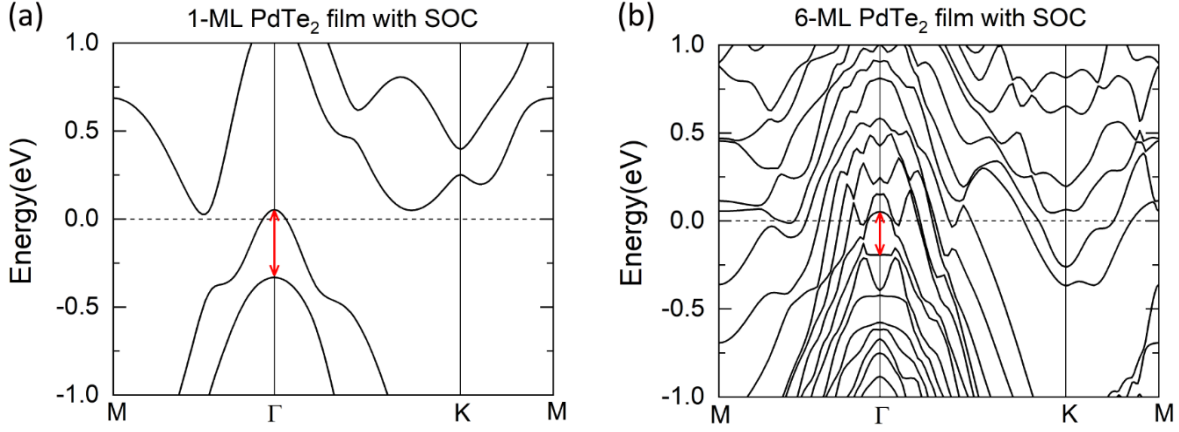

Figure S6. The band structures of 1-ML (a) and 6-ML (b) PdTe$_2$ films with SOC. The Zeeman-type spin splitting (marked as the red arrows) around the Γ point for 1-ML film is obviously larger than that of 6-ML film. The strong Zeeman-type SOC $\beta_{SO}$ can significantly suppress the spin splitting induced by the external in-plane magnetic field. Therefore, the effect of $\beta_{SO}$ can be roughly estimated by calculating the suppression of the spin splitting induced by the external in-plane magnetic field, which decreases rapidly with increasing momentum k as shown in Figure 2c. The superconductivity from a Fermi pocket at large k would be easily affected by the in-plane critical field due to a relatively small $\beta_{SO}$. As a result, the type-II Ising superconductivity in PdTe$_2$ thin films mainly comes from the small Fermi pocket around the Γ point. In theory, the type-II Ising superconductivity[9] in the PdTe$_2$ thin films is thickness dependent due to the following reasons: (1) The Zeeman-type SOC around the Γ point becomes larger for thinner films. For instance, the Zeeman-type spin splitting of 1-ML PdTe$_2$ film is obviously larger than that of 6-ML PdTe$_2$ film. The larger Zeeman-type SOC can result in a larger in-plane critical field for thinner films; (2) The Fermi level varies with different film thickness and affect the momentum k of the electrons. The Zeeman-type SOC is largely k-dependent, and hence is thickness dependent due to the change of the Fermi level.

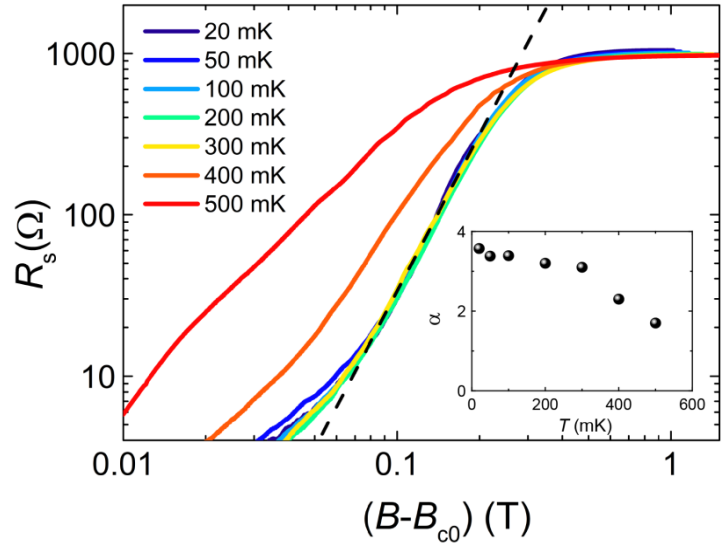

Figure S7. An alternative explanation of the anomalous metallic state is the Bose metal theory, which describes a gapless, nonsuperfluid state in the zero temperature limit [10-13]. In the framework of this theory, the small residual resistance at finite magnetic fields results from uncondensed Cooper pairs and vortices and the sheet resistance of the metallic state is proportional to $(B - B_{c0})^{2\nu}$, where $B_{c0}$ is the critical field of superconductor to Bose metal transition and $\nu$ is the exponent of the superfluid correlation length[11]. The magnetoresistance $R_s(B - B_{c0})$ curves of 4-ML PbTe$_2$ film is plotted in the double logarithmic scale at various temperatures from 20 mK to 500 mK. The black dashed line indicates the slope of the $R_s(B - B_{c0})$ curves in the low temperature regime. Here, $B_{c0}$ is defined as the critical field of zero resistance within the measurement resolution. The isotherms below 300 mK collapse to a single curve with the slope $\alpha = 2\nu$ slightly larger than 3. The inset shows the slope of the magnetoresistance as a function of temperature, which becomes relatively stable at ultralow temperatures, yielding a critical exponent $\nu \approx 1.72$.

Table S1 The fitting parameters of Eq. S1 for the perpendicular upper critical field of 4-ML and 6-ML PdTe2 films.

| Thickness | $\lambda_{11}$ | $\lambda_{12}$ | $\lambda_{22}$ | $D_1$ | $D_2$ | $T_c$/K |
|---|---|---|---|---|---|---|
| 4 ML | 30.7 | 19.8 | 13.9 | 1.21 | 0.269 | 0.704 |
| 6 ML | 2.74 | 9.35 | 29.2 | 29.0 | 0.584 | 0.737 |

Table S2 The fitting parameters of Eq. S4 for the in-plane critical field of 4-ML and 6-ML PdTe$_2$ films.

| Thickness | $\lambda_{11}$ | $\lambda_{12}$ | $\lambda_{22}$ | $\widetilde{\beta_{SO1}}$/ meV | $\widetilde{\beta_{SO2}}$/meV | $T_c$/K |
|---|---|---|---|---|---|---|
| 4 ML | 30.7 | 19.8 | 13.9 | 1.84 | 1.89 | 0.701 |
| 6 ML | 2.74 | 9.35 | 29.2 | 0.508 | 1.26 | 0.736 |

Table S3 The fitting parameters of Eq. 1 for the magnetoresistance of 4-ML PdTe$_2$ films from 20 to 100 mK.

| $T$/mK | $C$ | $R_n$/$\Omega$ | $B_{c2}^{\perp}$/T |
|---|---|---|---|
| 20 | 2.85 | 874.5 | 0.721 |
| 50 | 2.34 | 874.5 | 0.685 |
| 100 | 1.96 | 874.5 | 0.597 |